\documentclass[pra,superscriptaddress,twocolumn,amsmath,amssymb]{revtex4}% Physical Review A:twocolumn

\usepackage{graphicx}% Include figure files
\usepackage{bm}% bold math

\begin{document}

\title{Davies ENDOR revisited: \\ Enhanced sensitivity and nuclear spin relaxation}

\author{Alexei~M.~Tyryshkin}
\email{atyryshk@princeton.edu}
 \affiliation{Department of Electrical
Engineering, Princeton University, Princeton, NJ 08544, USA}

\author{John~J.~L.~Morton}
\affiliation{Department of Materials, Oxford University, Oxford
OX1 3PH, United Kingdom} \affiliation{Clarendon Laboratory,
Department of Physics, Oxford University, Oxford OX1 3PU, United
Kingdom}

\author{Arzhang~Ardavan}
\affiliation{Clarendon Laboratory, Department of Physics, Oxford
University, Oxford OX1 3PU, United Kingdom}

\author{S.~A.~Lyon}
\affiliation{Department of Electrical Engineering, Princeton
University, Princeton, NJ 08544, USA}

\date{\today}
% It is always \today, today, but any date may be explicitly specified

\begin{abstract}
Over the past 50 years, electron-nuclear double resonance (ENDOR)
has become a fairly ubiquitous spectroscopic technique, allowing
the study of spin transitions for nuclei which are coupled to
electron spins. However, the low spin number sensitivity of the
technique continues to pose serious limitations. Here we
demonstrate that signal intensity in a pulsed Davies ENDOR
experiment depends strongly on the nuclear relaxation time
T$_{1n}$, and can be severely reduced for long T$_{1n}$. We
suggest a development of the original Davies ENDOR sequence that
overcomes this limitation, thus offering dramatically enhanced
signal intensity and spectral resolution. Finally, we observe that
the sensitivity of the original Davies method to T$_{1n}$ can be
exploited to measure nuclear relaxation, as we demonstrate for
phosphorous donors in silicon and for endohedral fullerenes
N@C$_{60}$ in CS$_2$.

\end{abstract}

%\pacs{76.70.Dx, 76.60.-k, 71.55.Cn}

\maketitle

\section{Introduction}
Electron-nuclear double resonance (ENDOR) belongs to a powerful
family of polarization transfer spectroscopic methods and permits
the measurement of small energy (nuclear spin) transitions at the
much enhanced sensitivity of higher energy (electron spin)
transitions~\cite{feher56}. ENDOR is thus an alternative to NMR
methods, with the benefits of improved spin-number sensitivity and
a specific focus on NMR transitions of nuclei coupled to
paramagnetic species (reviewed in
Refs~\cite{kevan76,schweiger01}).

In an ENDOR experiment, the intensity of an electron paramagnetic
resonance (EPR) signal (e.g.~an absorption signal in continuous
wave EPR, or a spin echo signal in pulsed EPR) is monitored while
strong RF irradiation is applied to excite nuclear spin
transitions of the nuclei that are coupled to the electron spin.
Although the EPR signal may be strong, the RF-induced changes are
often rather weak and therefore it is quite common to find the
ENDOR signal to constitute only a few percent of the total EPR
signal intensity. Many different ENDOR schemes have been developed
to improve sensitivity and spectral resolution of the ENDOR signal
and to aid in analysis of congested ENDOR
spectra~\cite{kevan76,schweiger01,gemperle91}. However, low
visibility of the ENDOR signal remains a common problem to all
known ENDOR schemes, and long signal averaging (e.g.~hours to
days) is often required to observe the ENDOR spectrum at adequate
spectral signal/noise.

A low efficiency in spin polarization transfer (and thus low
intensity of the ENDOR response) is inherent to continuous wave
ENDOR experiments, which depend critically on accurate balancing
of the microwave and RF powers applied to saturate the electron
and nuclear spin transitions, and various spin relaxation times
within the coupled electron-nuclear spin system, including the
electron and nuclear spin-lattice relaxation times, T$_{1e}$ and
T$_{1n}$, and also the cross-relaxation (flip-flop) times,
T$_{1\rm{x}}$~\cite{dalton72}. The ENDOR signal is measured as a
partial de-saturation of the saturated EPR signal and generally
constitutes a small fraction of the full EPR signal
intensity~\cite{kevan76}. Since spin relaxation times are highly
temperature dependent, balancing these factors to obtain a maximal
ENDOR response is usually only possible within a narrow
temperature range.

Pulsed ENDOR provides many improvements over the continuous wave
ENDOR methods~\cite{gemperle91,schweiger01} and most importantly
eliminates the dependence on spin relaxation effects by performing
the experiment on a time scale which is short compared to the spin
relaxation times. Furthermore, combining microwave and RF pulses
enables 100$\%$ transfer of spin polarization, and therefore the
pulsed ENDOR response can in principle approach a 100$\%$
visibility (we define the ENDOR visibility as change in the echo
signal intensity induced by the RF pulse, normalized to the echo
intensity in the absence of the pulse~\cite{schweiger01,epel01}).
In practice, the situation is far from perfect and it is common to
observe a pulsed ENDOR response of the level of a few percent,
comparable to continuous wave ENDOR. In this paper we discuss the
limitations of the pulsed ENDOR method, and specifically Davies
ENDOR~\cite{davies74}. We suggest a modification to the pulse
sequence which dramatically enhances the signal/noise and can also
improve spectral resolution. We also show how traditional Davies
ENDOR may be used to perform a measurement of the nuclear
relaxation time, T$_{1n}$. While not discussed in this manuscript,
a similar modification is also applicable to Mims ENDOR
method~\cite{mims65}.

\section{Materials and Methods}
We demonstrate the new ENDOR techniques using two samples:
phosphorus $^{31}$P donors in silicon, and endohedral fullerenes
$^{14}$N@C$_{60}$ (also known as \emph{i}-NC$_{60}$) in CS$_2$
solvent. Silicon samples were epitaxial layers of
isotopically-purified $^{28}$Si (a residual $^{29}$Si
concentration of $\sim 800$~ppm as determined by secondary ion
mass spectrometry~\cite{itoh04}) grown on p-type natural silicon
(Isonics). The epi-layers were 10~$\mu$m thick and doped with
phosphorus at $1.6\cdot 10^{16}$ P/cm$^{3}$. Thirteen silicon
pieces (each of area 9$\times$3~mm$^2$) were stacked together to
form one EPR sample. This sample is referred as $^{28}$Si:P in the
text.

N@C$_{60}$ consists of an isolated nitrogen atom in the
$^4$S$_{3/2}$ electronic state incarcerated in a C$_{60}$
fullerene cage. Our production and subsequent purification of
N@C$_{60}$ is described elsewhere~\cite{mito}. High-purity
N@C$_{60}$ powder was dissolved in CS$_{2}$ to a final
concentration of 10$^{15}$/cm$^3$, freeze-pumped to remove oxygen,
and finally sealed in a quartz tube. Samples were 0.7~cm long, and
contained approximately $5\cdot 10^{13}$ N@C$_{60}$ molecules.

Both $^{28}$Si:P and N@C$_{60}$ can be described by a similar
isotropic spin Hamiltonian (in angular frequency units):
\begin{equation}\label{Hamiltonian}
\mathcal{H}_0=\omega_e S_z - \omega_I I_z + a \!\cdot\! \vec{S}
\!\cdot\! \vec{I},
\end{equation}
where $\omega_e=g\beta B_0/\hbar$ and $\omega_I=g_I\beta_n
B_0/\hbar$ are the electron and nuclear Zeeman frequencies, $g$
and $g_I$ are the electron and nuclear g-factors, $\beta$ and
$\beta_n$ are the Bohr and nuclear magnetons, $\hbar$ is Planck's
constant and $B_0$ is the magnetic field applied along $z$-axis in
the laboratory frame. In the case of $^{28}$Si:P, the electron
spin S=1/2 (g-factor = 1.9987) is coupled to the nuclear spin
I=1/2 of $^{31}$P through a hyperfine coupling $a=117$~MHz (or
4.19~mT)~\cite{fletcher54,feher59}. The X-band EPR signal of
$^{28}$Si:P consists of two lines (one for each nuclear spin
projection $M_I = \pm 1/2$). Our ENDOR measurements were performed
at the high-field line of the EPR doublet corresponding to
$M_I=-1/2$. In the case of N@C$_{60}$, the electron has a high
spin S=3/2 (g-factor = 2.0036) that is coupled to a nuclear spin
I=1 of $^{14}$N through an isotropic hyperfine coupling
$a=15.7$~MHz (or 0.56~mT)~\cite{murphy96}. The N@C$_{60}$ signal
comprises three lines and our ENDOR experiments were performed on
the central line ($M_I=0$) of the EPR triplet.

Pulsed EPR experiments were performed using an X-band Bruker EPR
spectrometer (Elexsys 580) equipped with a low temperature
helium-flow cryostat (Oxford CF935). The temperature was
controlled with a precision greater than $0.05$~K using calibrated
temperature sensors (Lakeshore Cernox CX-1050-SD) and an Oxford
ITC503 temperature controller. This precision was needed because
of the strong temperature dependence of the electron spin
relaxation times in the silicon samples (T$_{1e}$ varies by five
orders of magnitude between 7~K and 20~K)~\cite{alexeisi}.
Microwave pulses for $\pi$/2 and $\pi$ rotations of the electron
spin were set to 32 and 64~ns for the $^{28}$Si:P sample, and to
56 and 112~ns for the N@C$_{60}$ sample, respectively. In each
case the excitation bandwidth of the microwave pulses was greater
than the EPR spectral linewidth (e.g. 200~kHz for
$^{28}$Si:P~\cite{alexeisi}, and 8.4~kHz for
N@C$_{60}$~\cite{eseem05}) and therefore full excitation of the
signal was achieved. RF pulses of 20-50~$\mu$s were used for $\pi$
rotations of the $^{31}$P nuclear spins in $^{28}$Si:P and the
$^{14}$N nuclear spins in N@C$_{60}$.

\section{Standard Davies ENDOR Sequence}
\begin{figure}[t] \centerline
{\includegraphics[width=3.5in]{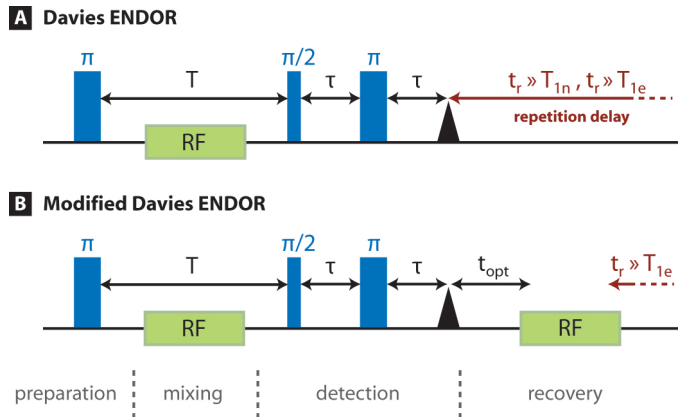}} \caption{Pulse
sequences for Davies ENDOR experiments. (A) The traditional Davies
experiment requires long recovery time t$_r \gg$ T$_{1e}$ and t$_r
\gg$ T$_{1n}$ to allow the spin system to fully recover to a thermal
equilibrium before the experiment can be repeated (e.g.~for signal
averaging). (B) An additional RF pulse applied after echo detection
helps the spin system to recover to a thermal equilibrium in a much
shorter time limited only by T$_{1e}$. Thus, signal averaging can be
performed at a much faster rate and an enhanced signal/noise can be
achieved in a shorter experimental time. t$_{\rm{opt}}$ represents
an optional delay of several
T$_{1e}$ which can be inserted for a secondary %(about 30\%)
improvement in signal/noise and to avoid overlapping with electron
spin coherences in case of long T$_{2e}$.} \label{fig1}
\end{figure}

\begin{figure*}[t] \centerline
{\includegraphics[width=7in]{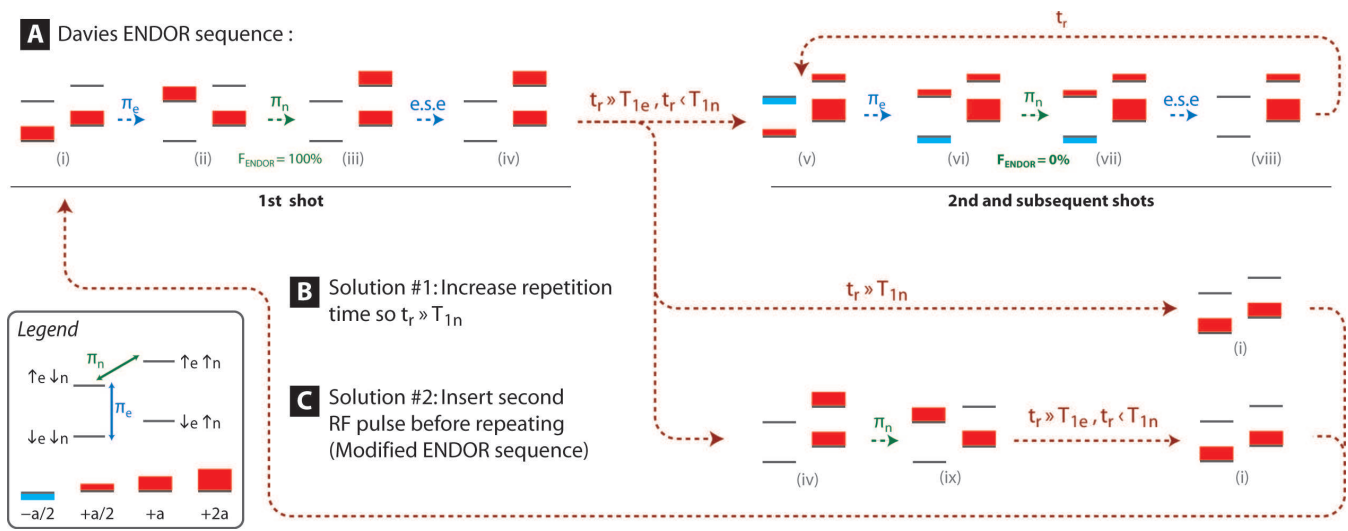}} \caption{Evolution of
spin state populations during the Davies ENDOR pulse sequence, for a
coupled electron $S=1/2$ and nucleus $I=1/2$. Legend shows an energy
level diagram and also electron ($\pi_e$) and nuclear ($\pi_n$) spin
transitions excited by selective microwave and RF pulses,
respectively. State populations are shown using colored bars (see
legend) in a high-temperature approximation ($a= g\mu_B B/kT \ll
1$), ignoring the identity component and also the small thermal
nuclear polarization. $g$ is the electron g factor, $\mu_B$ the Bohr
magneton and $B$ the applied magnetic field. The ENDOR visibility,
F$_{ENDOR}$, is measured as change in the electron spin polarization
which occurs between (ii) and (iii), or (vi) and (vii), caused by
the RF pulse. After the 2-pulse electron spin echo measurement
(e.s.e.), there is a long delay, t$_r$, before the experiment is
repeated. \textbf{(A)} In a typical experiment, T$_{1n}$,T$_{1x}
\gg$ t$_r \gg$ T$_{1e}$, and so only the electron spin has time to
relax --- (iv) relaxes to (v), not to thermal equilibrium (i). The
second and all subsequent experiments start from this new initial
state (v). The RF pulse produces no changes to the state population,
(vi) = (vii), and so the ENDOR signal is pessimal ($\sim 0\%$).
\textbf{(B)} One solution is to increase the repetition time so that
t$_r \gg$ T$_{1n}$,T$_{1x}$, although this can lead to a very slow
acquisition times. \textbf{(C)} A better solution is to apply an RF
pulse ($\pi_n$) after the electron spin echo formation and
detection. This additional RF pulse allows a faster return of the
spin system to a thermal equilibrium, e.g.~after several T$_{1e}$,
irrespective of T$_{1n}$ and T$_{1x}$.} \label{fig2}
\end{figure*}

Figure~\ref{fig1}A shows a schematic of the Davies ENDOR
sequence~\cite{davies74}, while Figure~\ref{fig2}A shows the
evolution of the spin state populations during the sequence (for
illustration purposes we consider a simple system of coupled
electron S=1/2 and nuclear I=1/2 spins, however the same
consideration is applicable to an arbitrary spin system). In the
\emph{preparation} step of the pulse sequence, a selective
microwave $\pi$ pulse is applied to one of the electron spin
transitions to transfer the initial thermal polarization (i) of
the electron spin to the nuclear spin polarization (ii). In the
\emph{mixing} step a resonant RF pulse on the nuclear spin further
disturbs the electron polarization to produce (iii), which can be
\emph{detected} using a two-pulse (Hahn) echo pulse
sequence~\cite{schweiger01}. A side result of the detection
sequence is to equalize populations of the resonant electron spin
states (iv). There then follows a delay, $t_r$, before the
experiment is repeated (e.g.~for signal averaging). Analysis of
this \emph{recovery} period has hitherto been limited (although
the effect of $t_r$ has been discussed with respect to ENDOR
lineshape~\cite{epel01} and stochastic ENDOR
acquisition~\cite{epel03}), yet it is this recovery period which
is crucial in optimizing the sequence sensitivity.

Nuclear spin relaxation times T$_{1n}$ (and also cross-relaxation
times T$_{1x}$) are usually very long, ranging from many seconds
to several hours, while electron spin relaxation times T$_{1e}$
are much shorter, typically in the range of microseconds to
milliseconds. In a typical EPR experiment,  t$_r$ is chosen to be
several T$_{1e}$ (i.e.~long enough for the electron spin to fully
relax, but short enough to perform the experiment in a reasonable
time). Thus, in practice it is generally the case that T$_{1n}
\gg$ t$_r \gg$ T$_{1e}$, i.e.~t$_r$ is short on the time scale of
T$_{1n}$ while long on the time scale of T$_{1e}$.
%This choice is determined by a desire to accumulate the signal at the highest possible repetition
%rate (not being limited by excessively long T$_{1n}$) and thus to
%perform an entire experiment in a reasonable experimental time.
With this choice of t$_r$, during the recovery period only the
electron spin (and not the nuclear spin) has time to relax before
the next experiment starts. As shown in Figure~\ref{fig2}A, the
second and all subsequent shots of the experiment will start from
initial state (v), and not from the thermal equilibrium (i). While
the first shot yields a 100$\%$ ENDOR visibility, subsequent
passes give strongly suppressed ENDOR signals. Upon signal
summation over a number of successive shots, the overall ENDOR
response is strongly diminished from its maximal intensity and
fails to achieve the theoretical 100$\%$ by a considerable margin.

One obvious solution to overcoming this limitation is to increase
the delay time $t_r$ so that it is long compared to the nuclear
spin relaxation time T$_{1n}$  (Figure~\ref{fig2}B). In other
words, t$_r \gg$ (T$_{1n}$, T$_{1x}$)$\gg$ T$_{1e}$, so that the
entire spin system (including electron and nuclear spins) has
sufficient time between successive experiments to fully relax to
thermal equilibrium. However, this can make the duration of an
experiment very long, and the advantage of an enhanced per-shot
sensitivity becomes less significant. From calculations provided
in the Appendix, it can be seen that an optimal trade-off between
signal/noise and experimental time is found at t$_r \approx
5/4$T$_{1n}$.

A better solution to this problem involves a modification of the
original Davies ENDOR sequence which removes the requirement for
t$_r$ to be greater than T$_{1n}$, permitting enhanced
signal/noise at much higher experimental repetition rates, limited
only by T$_{1e}$.

\section{Modified Davies ENDOR Sequence}
Our modified Davies ENDOR sequence is shown in Figure~\ref{fig1}B.
An additional RF pulse is introduced at the end of the sequence,
after echo signal formation and detection. This second RF pulse is
applied at the same RF frequency as the first RF pulse and its
sole purpose is to re-mix the spin state populations in such a way
that the spin system relaxes to thermal equilibrium on the
T$_{1e}$ timescale, independent on T$_{1n}$. The effect of this
second RF pulse is illustrated in Figure~\ref{fig2}C. After echo
signal detection, the spin system is in state (iv) and the second
RF pulse converts it to (ix). This latter state then relaxes to
thermal equilibrium (i) within a short t$_r$ ($> 3$T$_{1e}$). In
this modified sequence each successive shot is identical
%(e.g. starts with the thermal equilibrium state (i))
and therefore adds the optimal ENDOR visibility to the accumulated
signal.

The discussion in Figure~\ref{fig2}C assumes an ideal $\pi$
rotation by the RF pulses. However, in experiment the RF pulse
rotation angle may differ from $\pi$, and such an imperfection in
either RF pulse will lead to a reduction in the ENDOR signal.
Errors in the first pulse have the same effect as in a standard
Davies ENDOR experiment, reducing the ENDOR signal by a factor
$(1-\cos\theta)/2$, where $\theta$ the actual rotation angle.
Errors in the second RF pulse (and also accumulated errors after
the first pulse) cause incomplete recovery of spin system back to
the thermal equilibrium state (i) at the end of each shot, thus
reducing visibility of the ENDOR signal in the successive shots.
The pulse rotation errors can arise from inhomogeneity of the RF
field in the resonator cavity (e.g. spins in different parts of
the sample are rotated by different angle) or from off-resonance
excitation of the nuclear spins (when the excitation bandwidth of
the RF pulses is small compared to total width of the
inhomogeneously-broadened ENDOR line). It is desirable to
eliminate (or at least partially compensate) some of these errors
in experiment.

We find that introducing a delay t$_{\rm{opt}}$, to allow the
electron spin to fully relax before applying the second RF pulse
(Figure~\ref{fig1}B), helps to counter the effect of rotation
errors.
%In fact, the sequence t$_{\rm{opt}}$ -- RF($\pi$) -- t$_{r}$,
%where both t$_{\rm{opt}}$ and t$_{r}>$~3T$_{1e}$, represents an
%example of an universal "recovery" sequence which leads the spin
%system back to its full equilibrium, starting from arbitrary
%electron-nuclear spin level population state.
In numerical simulations, using the approach developed in
ref.~\cite{epel01,bowman00} and taking into account electron and
nuclear spin relaxation times and also a finite excitation
bandwidth of the RF pulses, we observed that introducing
t$_{\rm{opt}} \gg T_{1e}$ produces about 30\% increase in the
ENDOR signal visibility (however, at cost of a slower acquisition
rate with repetition time t$_{\rm{opt}}+$~t$_{r}$).
%In the experimental demonstrations below, we used this protocol with t$_{\rm{opt}}=~$t$_{r}$.
In the following sections, we demonstrate the capabilities of this
modified Davies ENDOR sequence, using two examples of phosphorous
donors in silicon and N@C$_{60}$ in CS$_2$.

%\footnote{If it is desirable to detect a broader linewidth (i.e.~to boost signal intensity),
%it is beneficial to wait for the electron spin to relax (i.e.
%3T$_{1e}$) \emph{before} applying the second RF pulse, and then
%wait a further 3T$_{1e}$ before starting the next experiment. This
%process (3T$_{1e}$, RF-pulse, 3T$_{1e}$) can also be repeated to
%enhance sensitivity further, but each addition has rapidly
%diminishing effects.}

\section{Application of the modified Davies ENDOR}
\subsection{Improved Sensitivity}
\begin{figure}[t] \centerline
{\includegraphics[width=3in]{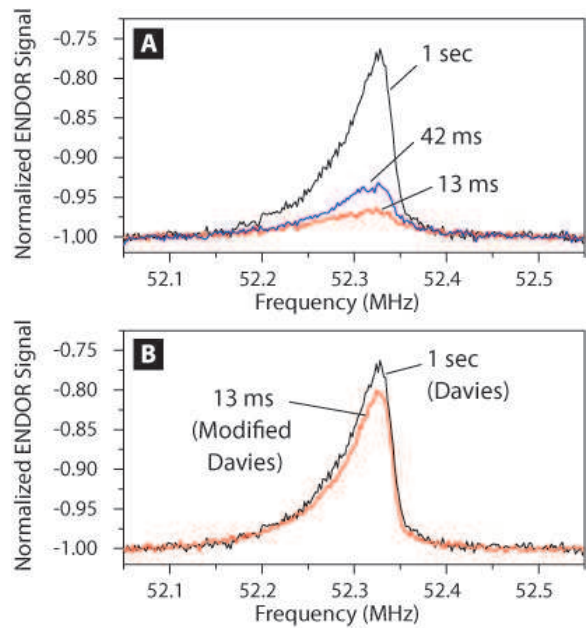}} \caption{Davies ENDOR
spectra for $^{28}$Si:P, showing the low-frequency $^{31}$P nuclear
transition line at 10~K. The spectra are normalized with respect to
the spin echo intensity with no RF pulse applied. (A) Three spectra
measured with the traditional Davies ENDOR pulse sequence (see
Figure~\ref{fig1}A) using different repetition times as labeled. The
same number of averages (n=20) was applied for each spectrum and
therefore the spectral acquisition times were approximately
proportional to the repetition times (i.e. 5000, 210 and 65 seconds,
respectively). (B) The spectrum measured with our modified Davies
pulse sequence (see Figure~\ref{fig1}B) using short repetition time
(13~ms) shows a comparable signal/noise to the spectrum measured
with a standard Davies pulse sequence at much longer repetition time
(1~s). t$_{\rm{opt}}$=6.5~ms was used in the modified Davies ENDOR
experiment.} \label{fig3}
\end{figure}

%Phosphorous donor in silicon ($^{28}$Si:P) has an electron spin $S=1/2$
%coupled to a nuclear spin $I=1/2$ of $^{31}$P and its energy level
%diagram is as shown in the legend to Figure~\ref{fig2}.
Figure~\ref{fig3}A shows the effect of experimental repetition
time, t$_r$, on the ENDOR visibility, using a standard Davies
ENDOR sequence applied to $^{28}$Si:P. Although t$_r$ is always
longer than the electron spin relaxation time (T$_{1e} = 1$~ms for
$^{28}$Si:P at 10~K~\cite{alexeisi}), increasing the repetition
time from 13~ms to 1 second improves the visibility by an order of
magnitude. As shown below, T$_{1n} = 288$~ms for the $^{31}$P
nuclear spin at 10~K, and therefore we observe that the ENDOR
signal visibility is weak ($\sim 2$\%) when t$_r = 13$~ms is
shorter than T$_{1n}$ but the visibility increases to a maximum
22\% when t$_r = 1$~s is longer than T$_{1n}$. The observed
maximal visibility 22\% does not reach a theoretical 100\% limit
because of the finite excitation bandwidth of the applied RF
pulses ($t_{RF}=50$~$\mu$s in these experiments) which is smaller
than total linewidth of the inhomogeneously-broadened $^{31}$P
ENDOR peak.

Through the use of the modified Davies ENDOR sequence proposed
above, the same order of signal enhancement is possible at the
faster 13~ms repetition time (e.g. at t$_r \ll$~T$_{1n}$), as
shown in Figure~\ref{fig3}B. This is an impressive improvement
indeed, considering that the acquisition time was almost 100 times
shorter in the modified Davies ENDOR experiment. The signal is
slightly smaller in the modified Davies ENDOR spectrum because of
the imperfect $\pi$ rotation of the recovery RF pulse (e.g. due to
inhomogeneity of the RF field as discussed above).
Figure~\ref{fig5}A shows a similar
signal enhancement effect for %a high-spin system of coupled $S=3/2$ and $I=1$ in
N@C$_{60}$.

\subsection{Improved Spectral Resolution}
Spectral resolution in a traditional Davies ENDOR experiment is
determined by the duration of the RF pulse inserted between the
preparation and detection microwave pulses (see
Figure~\ref{fig1}). The electron spin relaxation time T$_{1e}$
limits the maximum duration of this RF pulse, and in turn, the
achievable resolution in the ENDOR spectrum. However, there is no
such limitation on the duration of the second (recovery) RF pulse
in the modified Davies ENDOR sequence, as it is applied after the
electron spin echo detection. Thus, in the case where the duration
of the first RF pulse limits the ENDOR resolution, applying a
longer (and thus, more selective) second RF pulse can offer
substantially enhanced spectral resolution. In this scheme, the
first RF pulse is short and non-selectively excites a broad ENDOR
bandwidth, however the second RF pulse is longer and selects a
narrower bandwidth from the excited spectrum. Note that both RF
pulses  correspond to $\pi$ rotations, hence the power of the
second pulse must be reduced accordingly.

\begin{figure}[t] \centerline
{\includegraphics[width=3.2in]{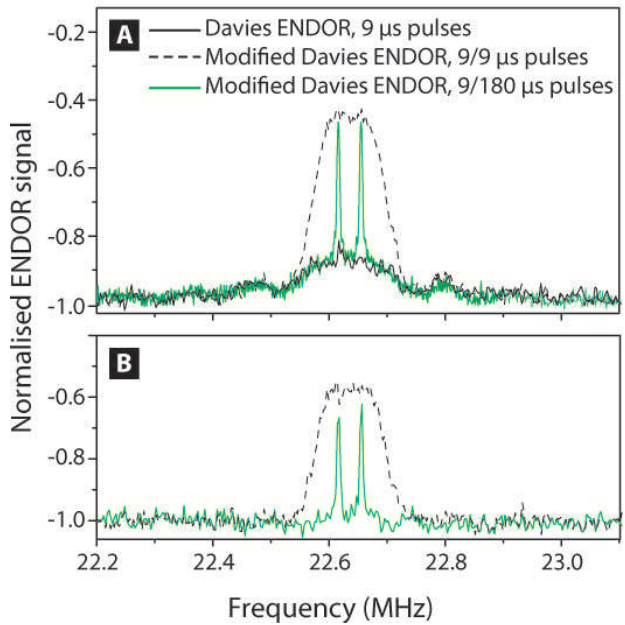}}
 \caption{Spectral resolution enhancement in the modified
Davies ENDOR experiment can be achieved by increasing the spectral
selectivity of the second (recovery) RF pulse, as illustrated for
the $M_S=-3/2$ ENDOR line of $^{14}$N@C$_{60}$, in CS$_2$ at
190~K. (A) A comparison of the traditional Davies ENDOR with a
9~$\mu$s RF pulse and the modified Davies ENDOR with an additional
9~$\mu$s recovery RF pulse demonstrates a significant enhancement
in signal intensity. If the second RF pulse is lengthened (to
180~$\mu$s in this case), the selectivity of the recovery pulse
increases and the enhanced component of the ENDOR line becomes
better resolved. (B) The oscillating background is identical in
all spectra in (A) and it can be removed from the modified Davies
ENDOR spectra by subtracting the spectrum obtained with a
traditional Davies ENDOR.} \label{fig5}
\end{figure}

Figure~\ref{fig5} illustrates this effect for N@C$_{60}$, in which
the intrinsic $^{14}$N~ENDOR lines are known to be very narrow
($<1$~kHz). %~\cite{weiden99}
Increasing the duration of the recovery RF pulse from
9~$\rm{\mu}$s to 180~$\rm{\mu}$s dramatically increases the
resolution and reveals two narrow lines, at no significant cost in
signal intensity or experiment duration. In Figure~\ref{fig5}B,
what appears to be a single broad line is thus resolved into two,
corresponding to two non-degenerate $\Delta M_I=1$ spin
transitions of $^{14}$N $I=1$ nuclear spin at electron spin
projection $M_S = -3/2$. We notice the presence of a broad
oscillating background in the modified Davies ENDOR spectra in
Figure~\ref{fig5}A. This background matches the signal detected
using a standard Davies ENDOR, where it is clearly seen to have a
recognizable \emph{sinc}-function shape (i.e.~its modulus) and
thus corresponds to the off-resonance excitation profile of the
first RF pulse. As shown in Figure~\ref{fig5}B, this background
signal can be successfully eliminated from the modified Davies
ENDOR spectra by subtracting the signal measured with a standard
Davies ENDOR.

%When the excitation bandwidth of the RF pulses is insufficient to excite
%the full nuclear line, RF $\pi$ pulses are not ideal. The performance of
%the revised ENDOR sequence is not uniform for all rotation angles, and
%drops off as the nuclear rotation angle deviates from $\pi$. This has
%the effect of making the detection bandwidth more selective for a given
%RF pulse duration, thus increasing the spectral resolution.

\section{Measuring Nuclear Spin Relaxation Times T$_{1n}$}
\begin{figure}[t] \centerline
{\includegraphics[width=3.5in]{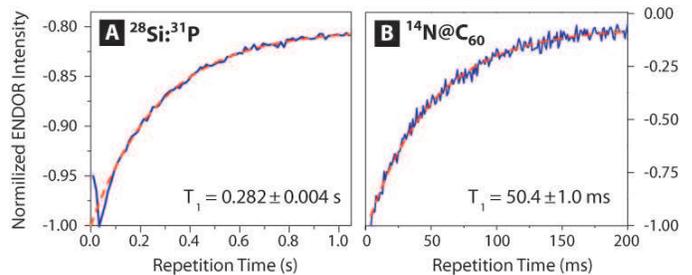}}
 \caption{Intensity of the traditional Davies ENDOR signal as a
function of repetition time measured for (A) $^{28}$Si:P
(52.33~MHz line) at 10~K, and (B) $^{14}$N@C$_{60}$ (22.626~MHz
line) in CS$_2$ at 190~K. An exponential fit (dashed line) yields
the respective nuclear spin relaxation time T$_{1\rm{n}}$. In (A)
the signal intensity at short times ($< 0.1$~s) deviates from the
exponential fit due to the transient effects arising from a finite
electron spin T$_{1e}$ time, as described in ref.~\cite{epel01}.}
\label{fig4}
\end{figure}

As already indicated in Figure~\ref{fig3}A, the signal intensity
in a traditional Davies ENDOR increases as the repetition time
t$_r$ is made longer, as compared to the nuclear spin relaxation
time T$_{1n}$. It is shown in the Appendix that, in case when
T$_{1n}\sim$~t$_r \gg$~T$_{1e}$, the ENDOR signal intensity varies
as:
\begin{equation}\label{ENDOR_tr_dependence}
I_{\rm{ENDOR}} \sim 1-\exp{\left(-t_r/T_{1n} \right)}.
\end{equation}
Thus, measuring the signal intensity in a traditional Davies ENDOR
as a function of t$_r$ yields a measure of T$_{1n}$, as
illustrated in Figure~\ref{fig4}A for $^{28}$Si:P and in
Figure~\ref{fig4}B for N@C$_{60}$. In both cases, T$_{1n}$ is
found to be much longer than T$_{1e}$ (cp. T$_{1n} = 280$~ms and
T$_{1e} = 1$~ms for $^{28}$Si:P at 10~K~\cite{alexeisi}, and
T$_{1n} = 50$~ms and T$_{1e} = 0.32$~ms for $^{14}$N@C$_{60}$ in
CS$_2$ at 190~K~\cite{relaxcs2}), as must be expected because
nuclear spins have a smaller magnetic moment and are therefore
less prone to fluctuating magnetic fields in the host environment.

Using Davies ENDOR to measure nuclear spin relaxation times,
T$_{1n}$ and T$_{2n}$, has been already proposed, however the
applicability of suggested pulse schemes has been limited to cases
where T$_{1n}$ (or~T$_{2n}$)~$<$~T$_{1e}$~\cite{hofer86,hofer94}.
Herein, we extend the method to (more common) cases where T$_{1n}$
is greater than T$_{1e}$.

\section{Conclusions}
We have shown that signal intensity in the traditional Davies
ENDOR  experiment is strongly dependent on the experimental
repetition time and that the addition of the second (recovery) RF
pulse at the end of the pulse sequence eliminates this dependence.
This modification to the Davies pulse sequence dramatically
enhances the signal/noise (allowing signal acquisition at much
faster rate without loss of the signal intensity), and can also
improve the spectral resolution. We also demonstrate that the
sensitivity of the Davies ENDOR to nuclear relaxation time can be
exploited to measure T$_{1n}$. The technique of adding an RF
recovery pulse after electron spin echo detection can be applied
to the general family of pulsed ENDOR experiments, in which a
non-thermal nuclear polarization is generated, including the
popular technique of Mims ENDOR~\cite{schweiger01,mims65}.

\section*{Acknowledgements}
We thank Kyriakos Porfyrakis for providing the N@C$_{60}$
material. We thank the Oxford-Princeton Link fund for support.
Work at Princeton was supported by the NSF International Office
through the Princeton MRSEC Grant No. DMR-0213706 and by the ARO
and ARDA under Contract No. DAAD19-02-1-0040. JJLM is supported by
St. John's College, Oxford. AA is supported by the Royal Society.

\section*{Appendix}
Here we describe how a compromise can be reached, using the
traditional Davies ENDOR, between maximal ENDOR `per-shot' signal
and overall experiment duration. The equations below show the
evolution of state populations --- a quantitative equivalent of
those shown in Figure~2 %\ref{fig2}
in the main text, with the difference that a partial nuclear
relaxation is considered during the repetition time $t_r$. Thus,
we assume T$_{1n}\sim$~t$_r \gg$~T$_{1e}$.
\begin{widetext}
\emph{Legend}:
\[
\left(%
\begin{array}{c}
  \uparrow_e \uparrow_n \\
  \uparrow_e \downarrow_n \\
  \downarrow_e \downarrow_n \\
  \downarrow_e \uparrow_n \\
\end{array}%
\right),~~~~a= g \beta_e
B/2kT~~~~~~~~~~~~~~~~~~~~~~~~~~~~~~~~~~~~~~~~~~~~~~~~~~~~~~~~~~~~~~~~~~~~~~~~~~~~~~~~~~~~~~~~~~~~~~~~~~~~~~~\]
1st shot:
\[
\left(%
\begin{array}{c}
  -a \\
  -a \\
  +a \\
  +a \\
\end{array}%
\right) \rightarrow \pi_e \rightarrow
\left(%
\begin{array}{c}
  -a \\
  +a \\
  -a \\
  +a \\
\end{array}%
\right) \rightarrow \pi_n \rightarrow
\left(%
\begin{array}{c}
  +a \\
  -a \\
  -a \\
  +a \\
\end{array}%
\right) \rightarrow e.s.e \rightarrow
\left(%
\begin{array}{c}
  +a \\
  -a \\
  -a \\
  +a \\
\end{array}%
\right) \rightarrow {\rm delay}~t_r \rightarrow
\left(%
\begin{array}{c}
  -a (1-\exp{\left(-t_r/T_{1n} \right)}) \\
  -a (1+\exp{\left(-t_r/T_{1n} \right)})\\
  +a (1-\exp{\left(-t_r/T_{1n} \right)})\\
  +a (1+\exp{\left(-t_r/T_{1n} \right)})\\
\end{array}%
\right)
\]
2nd and subsequent shots:
\[
\left(%
\begin{array}{c}
  -a (1-\exp{\left(-t_r/T_{1n} \right)})\\
  -a (1+\exp{\left(-t_r/T_{1n} \right)})\\
  +a (1-\exp{\left(-t_r/T_{1n} \right)})\\
  +a (1+\exp{\left(-t_r/T_{1n} \right)})\\
\end{array}%
\right) \rightarrow \pi_e \rightarrow
\left(%
\begin{array}{c}
  -a (1-\exp{\left(-t_r/T_{1n} \right)})\\
  +a (1-\exp{\left(-t_r/T_{1n} \right)})\\
  -a (1+\exp{\left(-t_r/T_{1n} \right)})\\
  +a (1+\exp{\left(-t_r/T_{1n} \right)})\\
\end{array}%
\right) \rightarrow \pi_n \rightarrow
\left(%
\begin{array}{c}
  +a (1-\exp{\left(-t_r/T_{1n} \right)})\\
  -a (1-\exp{\left(-t_r/T_{1n} \right)})\\
  -a (1+\exp{\left(-t_r/T_{1n} \right)})\\
  +a (1+\exp{\left(-t_r/T_{1n} \right)})\\
\end{array}%
\right) \rightarrow
\]
\[
\rightarrow e.s.e. \rightarrow
\left(%
\begin{array}{c}
  +a (1-\exp{\left(-t_r/T_{1n} \right)})\\
  -a \\
  -a \\
  +a (1+\exp{\left(-t_r/T_{1n} \right)})\\
\end{array}%
\right) \rightarrow {\rm delay}~t_r \rightarrow
\left(%
\begin{array}{c}
  -a (1-\exp{\left(-t_r/T_{1n} \right)}) \\
  -a (1+\exp{\left(-t_r/T_{1n} \right)})\\
  +a (1-\exp{\left(-t_r/T_{1n} \right)})\\
  +a (1+\exp{\left(-t_r/T_{1n} \right)})\\
\end{array}%
\right)
\]

\end{widetext}
The intensity of the ENDOR signal is therefore:
\[2a(1-\exp{\left(-t_r/T_{1n} \right)}).\]
As the signal-to-noise is proportional to the square root of the
number of samples, and thus to $\sqrt{1/t_r}$, we can define a
signal efficiency of:
\[2a(1-\exp{\left(-t_r/T_{1n} \right)})/\sqrt{t_r}.\]
This figure is maximized when $t_r \approx 1.25$~T$_{1n}$.

\bibliography{bib}

\begin{thebibliography}{20}
\expandafter\ifx\csname natexlab\endcsname\relax\def\natexlab#1{#1}\fi
\expandafter\ifx\csname bibnamefont\endcsname\relax
  \def\bibnamefont#1{#1}\fi
\expandafter\ifx\csname bibfnamefont\endcsname\relax
  \def\bibfnamefont#1{#1}\fi
\expandafter\ifx\csname citenamefont\endcsname\relax
  \def\citenamefont#1{#1}\fi
\expandafter\ifx\csname url\endcsname\relax
  \def\url#1{\texttt{#1}}\fi
\expandafter\ifx\csname urlprefix\endcsname\relax\def\urlprefix{URL }\fi
\providecommand{\bibinfo}[2]{#2}
\providecommand{\eprint}[2][]{\url{#2}}

\bibitem[{\citenamefont{Feher}(1956)}]{feher56}
\bibinfo{author}{\bibfnamefont{G.}~\bibnamefont{Feher}},
  \bibinfo{journal}{Phys. Rev.} \textbf{\bibinfo{volume}{103}},
  \bibinfo{pages}{834} (\bibinfo{year}{1956}).

\bibitem[{\citenamefont{Kevan and Kispert}(1976)}]{kevan76}
\bibinfo{author}{\bibfnamefont{L.}~\bibnamefont{Kevan}} \bibnamefont{and}
  \bibinfo{author}{\bibfnamefont{L.~D.} \bibnamefont{Kispert}},
  \emph{\bibinfo{title}{Electron spin double resonance spectroscopy}}
  (\bibinfo{publisher}{Wiley}, \bibinfo{address}{New York},
  \bibinfo{year}{1976}).

\bibitem[{\citenamefont{Schweiger and Jeschke}(2001)}]{schweiger01}
\bibinfo{author}{\bibfnamefont{A.}~\bibnamefont{Schweiger}} \bibnamefont{and}
  \bibinfo{author}{\bibfnamefont{G.}~\bibnamefont{Jeschke}},
  \emph{\bibinfo{title}{Principles of Pulse Electron Paramagnetic Resonance}}
  (\bibinfo{publisher}{Oxford University Press}, \bibinfo{address}{Oxford, UK ;
  New York}, \bibinfo{year}{2001}).

\bibitem[{\citenamefont{Gemperle and Schweiger}(1991)}]{gemperle91}
\bibinfo{author}{\bibfnamefont{C.}~\bibnamefont{Gemperle}} \bibnamefont{and}
  \bibinfo{author}{\bibfnamefont{A.}~\bibnamefont{Schweiger}},
  \bibinfo{journal}{Chem. Rev.} \textbf{\bibinfo{volume}{91}},
  \bibinfo{pages}{1481} (\bibinfo{year}{1991}).

\bibitem[{\citenamefont{Dalton and Kwiram}(1972)}]{dalton72}
\bibinfo{author}{\bibfnamefont{L.~R.} \bibnamefont{Dalton}} \bibnamefont{and}
  \bibinfo{author}{\bibfnamefont{A.~L.} \bibnamefont{Kwiram}},
  \bibinfo{journal}{J. Chem. Phys.} \textbf{\bibinfo{volume}{57}},
  \bibinfo{pages}{1132} (\bibinfo{year}{1972}).

\bibitem[{\citenamefont{Epel et~al.}(2001)\citenamefont{Epel, Poppl,
  Manikandan, Vega, and Goldfarb}}]{epel01}
\bibinfo{author}{\bibfnamefont{B.}~\bibnamefont{Epel}},
  \bibinfo{author}{\bibfnamefont{A.}~\bibnamefont{Poppl}},
  \bibinfo{author}{\bibfnamefont{P.}~\bibnamefont{Manikandan}},
  \bibinfo{author}{\bibfnamefont{S.}~\bibnamefont{Vega}}, \bibnamefont{and}
  \bibinfo{author}{\bibfnamefont{D.}~\bibnamefont{Goldfarb}},
  \bibinfo{journal}{J. Magn. Reson.} \textbf{\bibinfo{volume}{148}},
  \bibinfo{pages}{388} (\bibinfo{year}{2001}).

\bibitem[{\citenamefont{Davies}(1974)}]{davies74}
\bibinfo{author}{\bibfnamefont{E.~R.} \bibnamefont{Davies}},
  \bibinfo{journal}{Phys. Lett. A} \textbf{\bibinfo{volume}{47}},
  \bibinfo{pages}{1} (\bibinfo{year}{1974}).

\bibitem[{\citenamefont{Mims}(1965)}]{mims65}
\bibinfo{author}{\bibfnamefont{W.~B.} \bibnamefont{Mims}},
  \bibinfo{journal}{Proc. Roy. Soc. London} \textbf{\bibinfo{volume}{283}},
  \bibinfo{pages}{452} (\bibinfo{year}{1965}).

\bibitem[{\citenamefont{Itoh}(2004)}]{itoh04}
\bibinfo{author}{\bibfnamefont{K.~M.} \bibnamefont{Itoh}},
  \emph{\bibinfo{title}{private communication}} (\bibinfo{year}{2004}).

\bibitem[{\citenamefont{Kanai et~al.}(2004)\citenamefont{Kanai, Porfyrakis,
  Briggs, and Dennis}}]{mito}
\bibinfo{author}{\bibfnamefont{M.}~\bibnamefont{Kanai}},
  \bibinfo{author}{\bibfnamefont{K.}~\bibnamefont{Porfyrakis}},
  \bibinfo{author}{\bibfnamefont{G.~A.~D.} \bibnamefont{Briggs}},
  \bibnamefont{and} \bibinfo{author}{\bibfnamefont{T.~J.~S.}
  \bibnamefont{Dennis}}, \bibinfo{journal}{Chem. Commun.}
  \textbf{\bibinfo{volume}{2}}, \bibinfo{pages}{210} (\bibinfo{year}{2004}).

\bibitem[{\citenamefont{Fletcher et~al.}(1954)\citenamefont{Fletcher, Yager,
  Pearson, and Merritt}}]{fletcher54}
\bibinfo{author}{\bibfnamefont{R.~C.} \bibnamefont{Fletcher}},
  \bibinfo{author}{\bibfnamefont{W.~A.} \bibnamefont{Yager}},
  \bibinfo{author}{\bibfnamefont{G.~L.} \bibnamefont{Pearson}},
  \bibnamefont{and} \bibinfo{author}{\bibfnamefont{F.~R.}
  \bibnamefont{Merritt}}, \bibinfo{journal}{Phys. Rev.}
  \textbf{\bibinfo{volume}{95}}, \bibinfo{pages}{844} (\bibinfo{year}{1954}).

\bibitem[{\citenamefont{Feher}(1959)}]{feher59}
\bibinfo{author}{\bibfnamefont{G.}~\bibnamefont{Feher}},
  \bibinfo{journal}{Phys. Rev.} \textbf{\bibinfo{volume}{114}},
  \bibinfo{pages}{1219} (\bibinfo{year}{1959}).

\bibitem[{\citenamefont{Murphy et~al.}(1996)\citenamefont{Murphy, Pawlik,
  Weidinger, Hohne, Alcala, and Spaeth}}]{murphy96}
\bibinfo{author}{\bibfnamefont{T.~A.} \bibnamefont{Murphy}},
  \bibinfo{author}{\bibfnamefont{T.}~\bibnamefont{Pawlik}},
  \bibinfo{author}{\bibfnamefont{A.}~\bibnamefont{Weidinger}},
  \bibinfo{author}{\bibfnamefont{M.}~\bibnamefont{Hohne}},
  \bibinfo{author}{\bibfnamefont{R.}~\bibnamefont{Alcala}}, \bibnamefont{and}
  \bibinfo{author}{\bibfnamefont{J.-M.} \bibnamefont{Spaeth}},
  \bibinfo{journal}{Phys. Rev. Lett.} \textbf{\bibinfo{volume}{77}},
  \bibinfo{pages}{1075} (\bibinfo{year}{1996}).

\bibitem[{\citenamefont{Tyryshkin et~al.}(2003)\citenamefont{Tyryshkin, Lyon,
  Astashkin, and Raitsimring}}]{alexeisi}
\bibinfo{author}{\bibfnamefont{A.~M.} \bibnamefont{Tyryshkin}},
  \bibinfo{author}{\bibfnamefont{S.~A.} \bibnamefont{Lyon}},
  \bibinfo{author}{\bibfnamefont{A.~V.} \bibnamefont{Astashkin}},
  \bibnamefont{and} \bibinfo{author}{\bibfnamefont{A.~M.}
  \bibnamefont{Raitsimring}}, \bibinfo{journal}{Phys. Rev.B}
  \textbf{\bibinfo{volume}{68}}, \bibinfo{pages}{193207}
  (\bibinfo{year}{2003}).

\bibitem[{\citenamefont{Morton et~al.}(2005)\citenamefont{Morton, Tyryshkin,
  Ardavan, Porfyrakis, Lyon, and Briggs}}]{eseem05}
\bibinfo{author}{\bibfnamefont{J.~J.~L.} \bibnamefont{Morton}},
  \bibinfo{author}{\bibfnamefont{A.~M.} \bibnamefont{Tyryshkin}},
  \bibinfo{author}{\bibfnamefont{A.}~\bibnamefont{Ardavan}},
  \bibinfo{author}{\bibfnamefont{K.}~\bibnamefont{Porfyrakis}},
  \bibinfo{author}{\bibfnamefont{S.~A.} \bibnamefont{Lyon}}, \bibnamefont{and}
  \bibinfo{author}{\bibfnamefont{G.~A.~D.} \bibnamefont{Briggs}},
  \bibinfo{journal}{J. Chem. Phys.} \textbf{\bibinfo{volume}{122}},
  \bibinfo{pages}{174504} (\bibinfo{year}{2005}).

\bibitem[{\citenamefont{Epel et~al.}(2003)\citenamefont{Epel, Arieli, Baute,
  and Goldfarb}}]{epel03}
\bibinfo{author}{\bibfnamefont{B.}~\bibnamefont{Epel}},
  \bibinfo{author}{\bibfnamefont{D.}~\bibnamefont{Arieli}},
  \bibinfo{author}{\bibfnamefont{D.}~\bibnamefont{Baute}}, \bibnamefont{and}
  \bibinfo{author}{\bibfnamefont{D.}~\bibnamefont{Goldfarb}},
  \bibinfo{journal}{J. Magn. Reson.} \textbf{\bibinfo{volume}{164}},
  \bibinfo{pages}{78} (\bibinfo{year}{2003}).

\bibitem[{\citenamefont{Bowman and Tyryshkin}(2000)}]{bowman00}
\bibinfo{author}{\bibfnamefont{M.~K.} \bibnamefont{Bowman}} \bibnamefont{and}
  \bibinfo{author}{\bibfnamefont{A.~M.} \bibnamefont{Tyryshkin}},
  \bibinfo{journal}{J. Magn. Reson.} \textbf{\bibinfo{volume}{144}},
  \bibinfo{pages}{74} (\bibinfo{year}{2000}).

\bibitem[{\citenamefont{Morton et~al.}(2006)\citenamefont{Morton, Tyryshkin,
  Ardavan, Porfyrakis, Lyon, and Briggs}}]{relaxcs2}
\bibinfo{author}{\bibfnamefont{J.~J.~L.} \bibnamefont{Morton}},
  \bibinfo{author}{\bibfnamefont{A.~M.} \bibnamefont{Tyryshkin}},
  \bibinfo{author}{\bibfnamefont{A.}~\bibnamefont{Ardavan}},
  \bibinfo{author}{\bibfnamefont{K.}~\bibnamefont{Porfyrakis}},
  \bibinfo{author}{\bibfnamefont{S.~A.} \bibnamefont{Lyon}}, \bibnamefont{and}
  \bibinfo{author}{\bibfnamefont{G.~A.~D.} \bibnamefont{Briggs}},
  \bibinfo{journal}{J. Chem. Phys.} \textbf{\bibinfo{volume}{124}},
  \bibinfo{pages}{014508} (\bibinfo{year}{2006}).

\bibitem[{\citenamefont{H\"{o}fer et~al.}(1986)\citenamefont{H\"{o}fer, Grupp,
  and Mehring}}]{hofer86}
\bibinfo{author}{\bibfnamefont{P.}~\bibnamefont{H\"{o}fer}},
  \bibinfo{author}{\bibfnamefont{A.}~\bibnamefont{Grupp}}, \bibnamefont{and}
  \bibinfo{author}{\bibfnamefont{M.}~\bibnamefont{Mehring}},
  \bibinfo{journal}{Phys. Rev. A} \textbf{\bibinfo{volume}{33}},
  \bibinfo{pages}{3519} (\bibinfo{year}{1986}).

\bibitem[{\citenamefont{H\"{o}fer}(1994)}]{hofer94}
\bibinfo{author}{\bibfnamefont{P.}~\bibnamefont{H\"{o}fer}}, in
  \emph{\bibinfo{booktitle}{36th Rocky Mountain Conference on Analitycal
  Chemistry}} (\bibinfo{address}{Denver, CO}, \bibinfo{year}{1994}), p.
  \bibinfo{pages}{103}.

\end{thebibliography}

\end{document}